\documentclass[12pt]{article}

\begin{document}

\noindent {\small CITUSC/00-015\hfill \hfill hep-th/0106013 \newline
NSF-ITP-01-54\hfill \bigskip }

{\vskip0.5cm}

\begin{center}
{\Large \textbf{U}}$_{\star }${\Large \textbf{(1,1) Noncommutative Gauge
Theory }}

{\Large \textbf{As The Foundation of 2T-Physics in Field Theory}}{\footnote{%
This research was partially supported by the US Department of Energy under
grant number DE-FG03-84ER40168, and by the National Science Foundation under
Grant No. PHY99-07949.}}

\bigskip

{\vskip0.5cm}

\textbf{Itzhak Bars}

{\vskip0.5cm}

\textsl{CIT-USC Center for Theoretical Physics \& Department of Physics}

\textsl{University of Southern California,\ Los Angeles, CA 90089-2535, USA}

{\vskip0.5cm}

\textbf{Abstract}
\end{center}

A very simple field theory in noncommutative phase space $(X^{M},P^{M})$ in $%
d+2$ dimensions, with a gauge symmetry based on noncommutative u$_{\star
}\left( 1,1\right) $, furnishes the foundation for the field theoretic
formulation of Two-Time Physics$.$ This leads to a remarkable unification of
several gauge principles in $d$ dimensions, including Maxwell, Einstein and
high spin gauge principles, packaged together into one of the simplest
fundamental gauge symmetries in noncommutative quantum phase space in $d+2$
dimensions. A gauge invariant action is constructed and its nonlinear
equations of motion are analyzed. Besides elegantly reproducing the first
quantized worldline theory with all background fields, the field theory
prescribes unique interactions among the gauge fields. A matrix version of
the theory, with a large $N$ limit, is also outlined.

\section{Introduction}

Two-Time Physics (2T-physics) \cite{survey2T}-\cite{NCSp} is a device that
makes manifest many hidden features of one-time physics (1T-physics). Until
recently, most of the understanding in 2T-physics was gained from studying
the worldline formalism. This revealed a $d+2$ dimensional holographic
origin of certain aspects of 1T-physics in $d$ dimensions, including, in
particular, higher dimensional hidden symmetries (conformal, and others) and
new sets of duality-type relations among 1T dynamical systems. While the
physical phenomena described by 1T or 2T-physics are the same, the
space-time point of view is different. The 2T-physics approach in $d+2$
dimensions offers a highly symmetric and unified version of the phenomena
described by 1T-physics in $d$ dimensions. As such, it raises deep questions
about the meaning of space-time.

A noncommutative field theory in phase space introduced recently \cite{NCSp}
confirmed the worldline as well as the configuration space field theory \cite
{field2T} results of 2T-physics, and suggested more far reaching insights.
In this paper the approach of \cite{NCSp} will be taken one step further by
showing that it originates from a fundamental gauge symmetry principle based
on noncommutative u$_{\star }\left( 1,1\right) .$ We will see that this
phase space symmetry concisely unifies many gauge principles that are
traditionally formulated in configuration space separately from each other,
including the Maxwell, Einstein and high spin gauge principles.

All new phenomena in 2T-physics in the worldline formulation can be traced
to the presence of an essential gauge symmetry: sp$\left( 2,R\right) $
acting on phase space $\left( X^{M},P_{M}\right) $ \cite{old2T}. The 2T
feature of space-time (i.e. $X^{M}$ with two timelike dimensions) is not an
input, it is an outcome of the sp$\left( 2,R\right) $ gauge symmetry. Yet
this symmetry is responsible for the effective reduction of the $d+2$
dimensional two-time phase space to (a collection of) $d$ dimensional phase
spaces with one-time. Each of the $d$ dimensional phase spaces
holographically captures the contents of the $d+2$ dimensional theory, but
they do so with holographic pictures that correspond to different 1T
dynamics (different 1T Hamiltonians).

In the space of ``all worldline theories'' for a spinless particle (i.e. all
possible background fields), there is a symmetry generated by all canonical
transformations \cite{highspin}. These transformations are above and beyond
the local sp$\left( 2,R\right) $ on the worldline. As observed in \cite{NCSp}%
, the gauging of sp$\left( 2,R\right) $ in field theory (as opposed to
worldline theory) gives rise to a local noncommutative u$_{\star }$(1)
symmetry in noncommutative phase space that is closely connected to the
general canonical transformations. In this paper, we will see that the local
sp$\left( 2,R\right) $ combines with the local u$\left( 1\right) $ to form a
non-Abelian gauge symmetry described by the noncommutative Lie algebra u$%
_{\star }\left( 1,1\right) $ that will form the basis for the gauge theory
introduced in this paper (following the notation in \cite{ncOn}, we use the
star symbol $\star $ in denoting noncommutative symmetry groups).

The u$_{\star }\left( 1,1\right) $ gauge principle completes the formalism
of \cite{NCSp} into an elegant and concise theory which beautifully
describes 2T-physics in field theory in $d+2$ dimensions, while resolving
some problems that remained open. The resulting theory has deep connections
to standard $d$ dimensional gauge theories, gravity and the theory of high
spin fields \cite{vasil}.

There is also a finite matrix formulation of the theory in terms of u$\left(
N,N\right) $ matrices, such that the $N\rightarrow \infty $ limit becomes
the u$_{\star }\left( 1,1\right) $ gauge theory.

\subsection{Symmetries in the worldline theory}

The local symmetries that will play a role in noncommutative field \ theory
make a partial appearance in the worldline formalism. Therefore, for a self
contained set of arguments, we start from basic considerations of the
worldline formalism of 2T-physics for a spinless scalar particle.

The spinless particle is described in phase space by $X^{M}\left( \tau
\right) ,P_{M}\left( \tau \right) ,$ interacting with all possible
background fields. It is convenient to use the notation $X_{1}^{M}\equiv
X^{M}$ and $X_{2M}\equiv P_{M},$ with $i=1,2$ referring to $X_{i}.$ We avoid
introducing a background metric in $D$ dimensions by defining $X_{1}^{M}$
with an upper index and $X_{2M}$ with a lower index, and never raise or
lower the $M$ indices in the general setup, in the definitions of gauge
symmetries, or the construction of an action. Thus, the formalism is
background independent and is not a priori committed to any particular
signature of space-time. The signature is later determined dynamically by
the equations of motion. The worldline action has the form \cite{emgrav}\cite
{highspin} 
\begin{equation}
I_{Q}=\int d\tau \left[ \dot{X}_{1}^{M}X_{2M}-{\frac{1}{2}}A^{ij}\left( \tau
\right) \,Q_{ij}(X_{1},X_{2})\right] ,  \label{worldline}
\end{equation}
where the symmetric $A_{ij}=A_{ji}$ denotes three Sp$\left( 2,R\right) $
gauge fields, and the symmetric $Q_{ij}=Q_{ji}$ are three sp$\left(
2,R\right) $ generators. An expansion of $Q_{ij}(X_{1},X_{2})$ in powers of $%
X_{2M}$ in some local domain, $Q_{ij}(X_{1},X_{2})=\sum_{s}\left(
f_{ij}\left( X_{1}\right) \right) ^{M_{1}\cdots M_{s}}X_{2M_{1}}\cdots
X_{2M_{s}},$ defines all the possible background fields $\left( f_{ij}\left(
X_{1}\right) \right) ^{M_{1}\cdots M_{s}}$ in configuration space. The local
sp$\left( 2,R\right) $ gauge transformations are 
\begin{equation}
\delta X_{1}^{M}=-\omega ^{ij}\left( \tau \right) \frac{\partial Q_{ij}}{%
\partial X_{2M}},\quad \delta X_{2M}=\omega ^{ij}\left( \tau \right) \frac{%
\partial Q_{ij}}{\partial X_{1}^{M}},\quad \delta A^{ij}=\dot{\omega}%
^{ij}\left( \tau \right) +\left[ A,\omega \left( \tau \right) \right] ^{ij}.
\end{equation}
The action $I_{Q}$ is gauge invariant, with local parameters $\omega
^{ij}\left( \tau \right) ,$ provided the $Q_{ij}\left( X_{1},X_{2}\right) $
satisfy the sp$\left( 2,R\right) $ Lie algebra under Poisson brackets. This
is equivalent to a set of differential equations that must be satisfied by
the background fields $\left( f_{ij}\left( X_{1}\right) \right)
^{M_{1}\cdots M_{s}}$ \cite{emgrav}\cite{highspin}. The simplest solution is
the free case denoted by $Q_{ij}=q_{ij}$ (no background fields, only the
flat metric $\eta _{MN}$) 
\begin{equation}
q_{ij}=X_{i}^{M}X_{j}^{N}\eta _{MN}:\quad q_{11}=X_{1}\cdot X_{1},\quad
q_{12}=X_{1}\cdot X_{2},\quad \ \ q_{22}=X_{2}\cdot X_{2}.  \label{qij}
\end{equation}

Beyond the local sp$\left( 2,R\right) $ above, if one considers the ``space
of all worldline theories'' of the type $I_{Q}$, there is a symmetry that
leaves the form of the action invariant \cite{highspin}. The symmetry can be
interpreted as acting in the space of all possible background fields $\left(
f_{ij}\left( X_{1}\right) \right) ^{M_{1}\cdots M_{s}}$ that obey the sp$%
\left( 2,R\right) $ closure conditions. The transformations are given by all
canonical transformations that act infinitesimally in the form 
\begin{equation}
\delta _{0}X_{1}^{M}=-\frac{\partial \omega _{0}\left( X_{1},X_{2}\right) }{%
\partial X_{2M}},\quad \delta _{0}X_{2M}=\frac{\partial \omega _{0}\left(
X_{1},X_{2}\right) }{\partial X_{1}^{M}},  \label{canon}
\end{equation}
for any $\omega _{0}\left( X_{1},X_{2}\right) .$ Then $\delta _{0}Q_{ij}$ is
derived from Eq.(\ref{canon}) and given by the Poisson brackets $\delta
_{0}Q_{ij}=\left\{ Q_{ij},\omega _{0}\right\} .$ Under such transformations
the term $\int \dot{X}_{1}^{M}X_{2M}$ is invariant, and the action $I_{Q}$
is mapped to $I_{\tilde{Q}}$ where $\tilde{Q}_{ij}\left( X_{1},X_{2}\right)
=Q_{ij}\left( \tilde{X}_{1},\tilde{X}_{2}\right) ,$ with $\tilde{X}%
_{i}=X_{i}+\delta _{0}X_{i}.$ The new action $I_{\tilde{Q}}$ is in the space
of all theories of the form $I_{Q}$ since, by virtue of canonical
transformations, the new $\tilde{Q}_{ij}$ satisfies the sp$\left( 2,R\right) 
$ algebra under Poisson brackets if the old $Q_{ij}$ does. By taking
advantage of these symmetries all possible $Q_{ij}\left( X_{1},X_{2}\right) $%
, i.e. all possible background fields have been determined up to canonical
transformations \cite{highspin}. The solution will be recapitulated later in
this paper in section (2.2).

\subsection{Field equations from first quantized theory}

Instead of using wavefunctions in configuration space $\psi \left(
X_{1}^{M}\right) $, the quantum theory can be formulated equivalently in
phase space, \`{a} la Weyl-Wigner-Moyal \cite{weyl}-\cite{moyal}, by using
distributions in phase space $\varphi \left( X_{1}^{M},X_{2M}\right) $. The
phase space approach is natural in 2T-physics, because the sp$\left(
2,R\right) $ \ as well as the canonical transformations $\omega _{0}\left(
X_{1},X_{2}\right) $ are phase space symmetries that would be cumbersome to
discuss (if not impossible) in configuration space. Therefore, we find it
beneficial to discuss first quantization in terms of fields in phase space.
Sometimes we will use the notation $X^{m}\equiv \left(
X_{1}^{M},X_{2M}\right) $ with a single index $m$ that takes $2(d+2)$
values. The fields in phase space will be functions of the form $A\left(
X^{m}\right) .$ Products of fields $A,B$ always involve the associative
noncommutative Moyal star product 
\begin{equation}
\left( A\star B\right) \left( X\right) =\left. \exp \left( \frac{i}{2}\theta
^{mn}\frac{\partial }{\partial X^{m}}\frac{\partial }{\partial \tilde{X}^{n}}%
\right) A\left( X\right) B\left( \tilde{X}\right) \right| _{X=\tilde{X}},
\end{equation}
where $\theta ^{mn}=\hbar \delta _{\,\,N}^{M}\,\varepsilon _{ij},$ with $%
i=1,2,$ and $\varepsilon _{ij}$ is the antisymmetric sp$\left( 2,R\right) $
invariant metric (note that we have not used any space-time metric in this
expression). The star commutator between any two fields is defined by $\left[
A,B\right] _{\star }\equiv A\star B-B\star A.$ The phase space coordinates
satisfy $\left[ X^{m},X^{n}\right] _{\star }=i\theta ^{mn},$ which is
equivalent to the Heisenberg algebra for $\left( X_{1}^{M},X_{2M}\right) .$

As shown in \cite{NCSp}, first quantization of the worldline theory of Eq.(%
\ref{worldline}) is described by the noncommutative field equations 
\begin{eqnarray}
\left[ Q_{ij},Q_{kl}\right] _{\star } &=&i\left( \varepsilon
_{jk}Q_{il}+\varepsilon _{ik}Q_{jl}+\varepsilon _{jl}Q_{ik}+\varepsilon
_{il}Q_{jk}\right) ,  \label{sp2} \\
Q_{ij}\star \varphi &=&0.  \label{matter}
\end{eqnarray}
Eq.(\ref{sp2}) is the quantum version of the sp$\left( 2,R\right) $
conditions required by the worldline theory. Its general solution was given
in \cite{highspin}\cite{NCSp}, and will be recapitulated in Eqs.(\ref{bkg1}-%
\ref{hologr2}) below. It describes Maxwell, Einstein and high spin \textit{%
background} gauge fields (i.e. no dynamics). Spinless matter is coupled to
these background gauge fields in Eq.(\ref{matter}). The general solution of
this equation is a superposition of a basis of fields $\varphi \left(
X_{1},X_{2}\right) =\sum_{nm}c_{m}^{\,\,n}\varphi _{n}^{\,\,m}\left(
X_{1},X_{2}\right) $ where \cite{NCSp} 
\begin{eqnarray}
\varphi _{n}^{\,\,m}\left( X_{1},X_{2}\right) &=&\int d^{D}Y\,\,\psi
_{n}\left( X_{1}\right) \star e^{-iY^{M}X_{2M}}\star \chi _{m}^{\ast }\left(
X_{1}\right)  \label{wignerr} \\
&=&\int d^{D}Y\,\,\,\psi _{n}\left( X_{1}-\frac{Y}{2}\right)
\,\,e^{-iY^{M}X_{2M}}\,\,\chi _{m}^{\ast }\left( X_{1}+\frac{Y}{2}\right) .
\end{eqnarray}
According toWeyl's correspondence, the $\varphi _{n}^{\,\,m}\left(
X_{1},X_{2}\right) $ are related to Hilbert space outer products $\varphi
_{n}^{\,\,m}\sim |\psi _{n}\rangle \langle \chi _{m}|$. The $\varphi $
equation (\ref{matter}) is equivalent to the sp$\left( 2,R\right) $ singlet
conditions in the Hilbert space, $Q_{ij}|\psi \rangle =0,$ whose solutions
form a complete set of physical states $\left\{ |\psi _{n}\rangle \right\} $
that are gauge invariant under sp$\left( 2,R\right) .$ The solution space $%
\left\{ |\psi _{n}\rangle \right\} $ is non-empty and is unitary only when
space-time has precisely two timelike dimensions, no less and no more \cite
{field2T}\cite{NCSp}. In particular, for the free theory (i.e. no
backgrounds other than the flat metric $\eta _{MN}$, thus $Q_{ij}\rightarrow
q_{ij}$), the $\left\{ |\psi _{n}\rangle \right\} $ form the basis for the
unitary singleton or doubleton representation of SO$\left( d,2\right) $%
\footnote{%
All the SO$\left( d,2\right) $ Casimir operators for the singleton/doubleton
representation are fixed, in particular, the quadratic Casimir is $%
C_{2}\left( SO\left( d,2\right) \right) =1-d^{2}/4$. There are many
holographic solutions of the $d+2$ dimensional differential equations $%
q_{ij}\psi _{n}\left( X_{1}\right) =0$ in the form of $d$ dimensional
fields, all of which realize the singleton/doubleton representation. One of
the holographic solutions is the Klein-Gordon field in $d$ dimensions which
forms a well known representation of the conformal group SO$\left(
d,2\right) .$ Another one is the Hydrogen atom in $d-1$ space dimensions,
another one is the scalar field in AdS$_{d},$ and still another one is the
scalar field in AdS$_{d-k}\times $S$^{k}$ for any $k<d-2,$ and more. They
all realize the same SO$\left( d,2\right) $ representation with the same
Casimir eigenvalues, but in different bases \cite{field2T}$.$} $.$ Unlike
the $\psi _{n}\left( X_{1}\right) ,$ the $\chi _{m}^{\ast }\left(
X_{1}\right) $ are not restricted by Eq.(\ref{matter}). Therefore, it is
reasonable to define $\varphi $ only up to noncommutative u$_{\star
}^{R}\left( 1\right) $ gauge transformations that act from the right $%
\varphi \rightarrow \varphi \star \exp _{\star }\left( -i\omega ^{R}\right) $%
, or to restrict it by an additional condition on $\varphi $ from the right
side. The $\varphi _{n}^{\,\,m}$ automatically satisfy the following closure
property under the triple product 
\begin{equation}
\varphi _{n_{1}}^{\,\,m_{1}}\star \left( \varphi ^{\dagger }\right)
_{m_{2}}^{\,\,n_{2}}\star \varphi _{n_{3}}^{\,\,m_{3}}=\delta
_{\,\,m_{2}}^{m_{1}}\,\delta _{\,\,n_{3}}^{n_{2}}\,\varphi
_{n_{1}}^{\,\,m_{3}},  \label{tripple}
\end{equation}
which follows just from the structure $|\psi _{n}\rangle \langle \chi _{m}|$
for orthonormalized states. By fixing a u$_{\star }^{R}\left( 1\right) $
gauge symmetry, $\varphi $ can be made Hermitian; in this case the set of $%
\left\{ \chi _{m}\right\} $ is the same as the $\left\{ \psi _{n}\right\} .$
Evidently, there are other choices for the gauge fixing of the right side of 
$\varphi $.

Eqs. (\ref{sp2},\ref{matter}) correctly represent quantum mechanically the
1T physics of a spinless particle in $d$ dimensions interacting with
background gauge fields, including the electromagnetic, gravitational and
high spin gauge fields \cite{NCSp}. Furthermore, the 2T physics formalism
unifies different types of 1T field theories in $d$ dimensions which
holographically represent the same $d+2$ dimensional equations, and
therefore, in principle it uncovers hidden symmetries and duality type
relations among them (this has been explicitly demonstrated in simple cases 
\cite{survey2T}).

Much of the work in \cite{NCSp} was devoted to developing the noncommutative
field theory formalism and the symmetry principles compatible with global
and local sp$\left( 2,R\right) $ symmetry. The goal was to find a field
theory, and appropriate gauge principles, from which the free Eqs.(\ref{sp2},%
\ref{matter}) would follow as classical field equations of motion, much in
the same way that the Klein-Gordon field theory arises from satisfying $\tau 
$-reparametrization constraints ($p^{2}=0$), or string field theory emerges
from satisfying Virasoro constraints, etc. This goal was partly accomplished
in \cite{NCSp}, but as we will explain, by only partially implementing the
full gauge principles described by u$_{\star }\left( 1,1\right) $.

In the rest of the paper we will complete the goal of \cite{NCSp} by
spelling out the gauge principles, and constructing an essentially unique
elegant action that results in Eqs.(\ref{sp2},\ref{matter}) as \textit{exact
background solutions} of nonlinear equations of motion. Expanding the full
equations of motion around any background solution provides consistent
interactions and propagation for the fluctuating gauge fields. Among other
nice features, this theory seems to provide an action principle for high
spin gauge fields.

\section{u$_{\star }$(1,1) gauge symmetry}

Global sp$\left( 2,R\right) $ transformations that treat $\left(
X_{1},X_{2}\right) $ as a doublet are generated by the free $%
q_{ij}=X_{i}\cdot X_{j}.$ A complex scalar field in phase space $\varphi
\left( X_{1},X_{2}\right) $ can transform as a left module, right module, or
diagonal module, as explained in \cite{NCSp}. For a left scalar module, the
global sp$\left( 2,R\right) $ transformation is $\delta ^{sp}\varphi
=-i\omega ^{ij}\left( q_{ij}\star \varphi \right) $ where $\omega ^{ij}$ are
global parameters. To turn sp$\left( 2,R\right) $ into a local symmetry, the
three $\omega ^{ij}$ are replaced by arbitrary functions. Then the Hermitian
combination $\omega _{0}=\frac{1}{2}\left( \omega ^{ij}\star
q_{ij}+q_{ij}\star \omega ^{ij}\right) $ acts from the left like a local
noncommutative phase transformation $\delta ^{0}\varphi =-i\omega _{0}\star
\varphi .$ Therefore, local sp$\left( 2,R\right) $ acting on a scalar field
from the left is closely related to a noncommutative local u$_{\star }\left(
1\right) .$ In \cite{NCSp} it is argued that this u$_{\star }\left( 1\right) 
$ acts on $Q_{ij}\left( X_{1},X_{2}\right) $ from both sides $\delta
^{0}Q_{ij}=-i\left[ \omega _{0},Q_{ij}\right] _{\star },$ therefore $\omega
_{0}\left( X_{1},X_{2}\right) $ is precisely the quantum version (all powers
of $\hbar $) of the canonical transformations, encountered in the worldline
formalism, as stated just following Eq.(\ref{canon}).

On a tensor field, global sp$\left( 2,R\right) $ acts both on its $\left(
X_{1}^{M},X_{2M}\right) $ dependence, as well as on its indices. For
example, for a doublet 
\begin{equation}
\delta _{global}^{sp}\varphi _{k}=\omega _{kl}\varphi ^{l}-i\omega
^{ij}\left( q_{ij}\star \varphi _{k}\right) .
\end{equation}
In turning these transformations into local transformations we find that we
must have independent local parameters $\omega _{ij}\left(
X_{1},X_{2}\right) $ and $\omega _{0}\left( X_{1},X_{2}\right) $ because
closure cannot be obtained with only the three parameters $\omega
_{ij}\left( X_{1},X_{2}\right) .$ In fact, there is no 3-parameter
noncommutative sp$\left( 2,R\right) $, instead there exists the local
four-parameter noncommutative u$_{\star }\left( 1,1\right) $ that has sp$%
\left( 2,R\right) =$ su$\left( 1,1\right) $ as a global subalgebra\footnote{%
For local parameters in noncommutative space, the commutator of two
transformations with $\omega _{ij}$ closes into a transformation that
involves both $\omega _{ij}$ and $\omega _{0}.$ The minimal noncommutative
algebra that includes sp$\left( 2,R\right) $ in the global limit is the
4-parameter sp$_{\star }\left( 2,R\right) $ \cite{ncOn}. This is a
subalgebra of the 4-parameter u$_{\star }\left( 1,1\right) $ and is obtained
from it by introducing a projection in the local space, such as the
interchange $X_{1}$ $\rightleftharpoons $ $X_{2},$ or mirror reflections $%
X_{2}\rightarrow -X_{2}$ as in \cite{ncOn}$.$ Thus, sp$_{\star }\left(
2,R\right) $ is embeded in u$_{\star }\left( 1,1\right) $ by demanding $%
\omega _{ij}\left( X_{1},X_{2}\right) $ to be symmetric functions and $%
\omega _{0}\left( X_{1},X_{2}\right) $ to be an antisymmetric function under
the projections. Closure is satisfied for the 4 projected functions. Thus,
in principle, sp$_{\star }\left( 2,R\right) $ would have been the minimal
local symmetry to turn global sp$\left( 2,R\right) $ into a local symmetry
in noncommutative space. However, as we will see, the simplest cubic action
that we will build for the gauge theory is not symmetric under the
parity-like projections. Therefore, the local symmetry appropriate for our
purposes is u$_{\star }\left( 1,1\right) $ rather than sp$_{\star }\left(
2,R\right) .$}. We can collect the 4 parameters in the form of a 2$\times $2
matrix, $\Omega _{ij}=\omega _{ij}+i\omega _{0}\varepsilon _{ij},$ whose
symmetric part $\omega _{ij}\left( X_{1},X_{2}\right) $ becomes sp$\left(
2,R\right) $ when it is global, while its antisymmetric part generates the
local subgroup u$_{\star }\left( 1\right) $ with local parameter $\omega
_{0}\left( X_{1},X_{2}\right) $. To act on the doublet one of the indices is
raised with the sp$\left( 2\right) $ metric $\varepsilon ^{ij},$ $\delta
\varphi _{k}=\Omega _{k}^{\,\,l}\star \varphi _{l}=\omega _{k}^{\,\,l}\star
\varphi _{l}-i\omega _{0}\star \varphi _{k},$ therefore in matrix form we
have 
\begin{equation}
\Omega _{k}^{\,\,l}=\omega _{k}^{\,\,l}-i\omega _{0}\delta
_{k}^{\,\,l}=\left( 
\begin{array}{cc}
\omega _{12}-i\omega _{0} & \omega _{22} \\ 
-\omega _{11} & -\omega _{12}-i\omega _{0}
\end{array}
\right)
\end{equation}
This matrix satisfies the following hermiticity conditions 
\begin{equation}
\Omega ^{\dagger }=\varepsilon \Omega \varepsilon .
\end{equation}
Such matrices close under matrix-star commutators to form u$_{\star }\left(
1,1\right) .$ It can be easily seen that, for closure under both matrix and
star products in commutators, $\omega _{ij}$ cannot be separated from the $%
\omega _{0}$ and hence they are both integral parts of the local symmetry.
The finite u$_{\star }\left( 1,1\right) $ group elements are given by
exponentiation (using star and matrix products) 
\begin{equation}
U=e_{\star }^{\Omega },\quad U^{-1}=\left( -\varepsilon \right) U^{\dagger
}\varepsilon =e_{\star }^{-\Omega },  \label{U}
\end{equation}

We can now consider the gauge fields. In \cite{NCSp} it was explained that $%
q_{ij}$ acting on $\varphi $ from the left defines a differential operator
that is appropriate for building the kinetic terms in the action. To turn
these differential operators into covariant differential operators, a gauge
potential $A_{ij}\left( X_{1},X_{2}\right) $ was introduced and added to the
differential operators when acting on $\varphi .$ Hence the covariant
derivatives are $Q_{ij}\left( X_{1},X_{2}\right) =q_{ij}+A_{ij}\left(
X_{1},X_{2}\right) $ acting from the left on $\varphi .$ These $Q_{ij}$ were
shown to play the same role as the sp$\left( 2,R\right) $ generators
encountered in the first quantized worldline theory. This was appropriate
for a scalar field, for which only u$_{\star }\left( 1\right) $ acts.
However, if we consider tensor fields, we must take covariant derivatives
with respect to u$_{\star }\left( 1,1\right) .$ Therefore we need to add
only one more gauge field or generator since u$_{\star }\left( 1,1\right) $
has 4 parameters. As we will see these will emerge from the following
considerations.

We introduce a 2$\times $2 matrix $\mathcal{J}_{ij}=J_{ij}+iJ_{0}\varepsilon
_{ij}$ that parallels the form of the parameters $\Omega _{ij}.$ There will
be a close relation between the fields $J_{ij}$ and $Q_{ij}$ as we will see
soon. When one of the indices is raised, the matrix takes the form 
\begin{equation}
\mathcal{J}_{i}^{\,\,\,\,\,j}=\left( 
\begin{array}{cc}
J_{12}-iJ_{0} & J_{22} \\ 
-J_{11} & -J_{12}-iJ_{0}
\end{array}
\right)  \label{J}
\end{equation}
The hermiticity of the fields $J_{ij},J_{0}$ is equivalent to the following u%
$_{\star }\left( 1,1\right) $ condition on this matrix 
\begin{equation}
\mathcal{J}^{\dagger }=\varepsilon \mathcal{J}\varepsilon .
\end{equation}
Local gauge transformations are defined by the matrix-star products in the
form 
\begin{equation}
\delta \mathcal{J}=\mathcal{J}\star \Omega -\Omega \star \mathcal{J},\quad
or\quad \mathcal{J}^{\prime }=U^{-1}\star \mathcal{J}\star U
\end{equation}
Then the matrix form and hermiticity of $\delta \mathcal{J}$ or $\mathcal{J}%
^{\prime }$ are consistent with the matrix form and hermiticity of $\mathcal{%
J}.$

Next we consider matter fields. For our purpose we will need to consider the
noncommutative group U$_{\star }^{L}\left( 1,1\right) \times $U$_{\star
}^{R}\left( 1,1\right) .$ Recall that we already had a hint that the matter
field admits independent gauge transformations on its left and right sides.
In this notation $\mathcal{J}$ transforms as the adjoint under U$_{\star
}^{L}\left( 1,1\right) $ and is a singlet under U$_{\star }^{R}\left(
1,1\right) ,$ thus it is in the $\left( 1,0\right) $ representation. For the
matter field we take the $\left( \frac{1}{2},\frac{1}{2}\right) $
representation given by a 2$\times 2$ complex matrix $\Phi _{i}^{\,\alpha
}\left( X_{1},X_{2}\right) .$ This field is equivalent to a complex
symmetric tensor $Z_{ij}$ and a complex scalar $\varphi ,$ both of which
were considered in \cite{NCSp}, but without realizing their U$_{\star
}^{L}\left( 1,1\right) \times $U$_{\star }^{R}\left( 1,1\right) $
classification. We define $\bar{\Phi}=\varepsilon \Phi ^{\dagger
}\varepsilon .$ The U$_{\star }^{L}\left( 1,1\right) \times $U$_{\star
}^{R}\left( 1,1\right) $ transformation rules for this field are 
\begin{equation}
\Phi ^{\prime }=U^{-1}\star \Phi \star W,\quad \bar{\Phi}^{\prime
}=W^{-1}\star \bar{\Phi}\star U.
\end{equation}
where $U\in $U$_{\star }^{L}\left( 1,1\right) $ and $W\in $U$_{\star
}^{R}\left( 1,1\right) .$

We now construct an action that will give the noncommutative field theory
equations (\ref{sp2},\ref{matter}) in a linearized approximation and
prescribe unique interactions in its full version. The action has a
resemblance to the Chern-Simons type action introduced in \cite{NCSp}, now
with an additional field, $J_{0},$ while the couplings among the fields obey
a higher gauge symmetry 
\begin{equation}
S_{\mathcal{J},\Phi }=\int d^{2D}X\,\,Tr\left( -\frac{i}{3}\mathcal{J}\star 
\mathcal{J}\star \mathcal{J}-\mathcal{J}\star \mathcal{J}\,\mathbf{+\,}i\bar{%
\Phi}\star \mathcal{J}\star \Phi -V_{\star }\left( \bar{\Phi}\star \Phi
\right) \right) .  \label{SZ}
\end{equation}
The invariance under the local U$_{\star }^{L}\left( 1,1\right) \times $U$%
_{\star }^{R}\left( 1,1\right) $ transformations is evident\footnote{%
More generally, the invariance under the local U$_{\star }^{R}\left(
1,1\right) $ could be broken by various terms in the potential $V_{\star
}\left( \bar{\Phi}\star \Phi \right) $ or by interactions of $\Phi $ with
additional fields from its right side.}. $V\left( u\right) $ is a potential
function with argument $u$=$\bar{\Phi}\star \Phi $. Although we will be able
to treat the most general potential function $V\left( u\right) $ in the
discussion below, to illustrate how the model works, it is sufficient to
consider the linear function $V\left( u\right) =au,$ which implies a
quadratic form in the field $\Phi $ 
\begin{equation}
V=a\,\bar{\Phi}\star \Phi .  \label{V}
\end{equation}
where $a$ is a constant.

The form of this action is unique as long as the maximum power of $\mathcal{J%
}$ is three. As we will see, when the maximum power of $\mathcal{J}$ is
cubic we will make the connection to the first quantized worldline theory.
We have not imposed any conditions on the powers of $\Phi $ or interactions
between $\mathcal{J},\Phi ,$ other than obeying the gauge symmetries. A
possible linear term in $\mathcal{J}$ can be eliminated by shifting $%
\mathcal{J}$ by a constant, while the relative coefficients in the action
are all absorbed into a renormalization of $\mathcal{J},\Phi $. A term of
the form $Tr\left( \mathcal{J}\star \mathcal{J}\star f\left( \Phi \star \bar{%
\Phi}\right) \right) $ that is allowed by the gauge symmetries can be
eliminated by shifting $\mathcal{J}\rightarrow \left( \mathcal{J}-\frac{1}{3}%
f\left( \Phi \star \bar{\Phi}\right) \right) .$ This changes the term $%
\mathbf{\,}i\bar{\Phi}\star \mathcal{J}\star \Phi $ by replacing it with
interactions of $\mathcal{J}$ with any function of $\bar{\Phi},\Phi $ that
preserves the gauge symmetries$.$ However, one can do field redefinitions to
define a new $\Phi $ so that the interaction with the linear $\mathcal{J}$
is rewritten as given, thus shifting all complications to the function $%
V_{\star }\left( \bar{\Phi}\star \Phi \right) $. Therefore, with the only
assumption being the cubic restriction on $\mathcal{J},$ this unique action
will explain the first quantized worldline theory, and will generalize it to
an interacting theory based purely on a gauge principle.

It can be checked that the action is hermitian thanks to the hermiticity
relations for $\mathcal{J},\Phi ,\bar{\Phi}$ and cyclicity under the trace
and integral signs. Hermiticity is also evident by evaluating the trace
explicitly\footnote{%
In this form we see that sp$_{\star }\left( 2,R\right) $ (as opposed to u$%
_{\star }\left( 1,1\right) $) cannot be used as the local symmetry, because
all cubic terms change sign under sp$_{\star }\left( 2,R\right) $'s
parity-like projections, $J_{ij}\left( X_{1},-X_{2}\right) =J_{ij}\left(
X_{1},X_{2}\right) $ and $J_{0}\left( X_{1},-X_{2}\right) =-J_{0}\left(
X_{1},X_{2}\right) $, with simultaneous interchange of factors in a star
product \cite{ncOn}. It is interesting to note that the action is invariant
if the parity properties are exactly the opposite signs than those required
by sp$_{\star }\left( 2,R\right) .$ However, such conditions could not be
imposed on $J$ because then they would not be compatible with gauge
transformations rules that are required to have a symmetric action.} 
\[
S_{\mathcal{J},\Phi }=\int d^{2D}X\left( 
\begin{array}{c}
iJ_{11}\star J_{12}\star J_{22}-iJ_{22}\star J_{12}\star J_{11}+\frac{2}{3}%
J_{0}\star J_{0}\star J_{0}+2J_{0}\star J_{0} \\ 
+\left( J_{11}\star J_{22}+J_{22}\star J_{11}-2J_{12}\star J_{12}\right)
\star \left( J_{0}+1\right) \mathbf{+}Tr\left( i\mathcal{J}\star \Phi \star 
\bar{\Phi}-V\right)
\end{array}
\right) . 
\]
The equations of motion are 
\begin{equation}
\mathcal{J}\star \Phi =-i\Phi \star V^{\prime },\quad \mathcal{J}\star 
\mathcal{J}\mathbf{-}2i\mathcal{J}\mathbf{-}\Phi \star \bar{\Phi}\mathbf{=}0.
\label{generaleq}
\end{equation}
where $V^{\prime }\left( u\right) =\partial V/\partial u$.

\subsection{Solution and link to worldline theory}

One can choose a gauge for the local U$_{\star }^{L}\left( 1,1\right) \times 
$U$_{\star }^{R}\left( 1,1\right) $ in which the 2$\times 2$ complex matrix $%
\Phi $ is proportional to the identity matrix 
\begin{equation}
\Phi _{i}^{\,\,\alpha }=\delta _{i}^{\,\,\alpha }\varphi \left(
X_{1},X_{2}\right) .
\end{equation}
Then $\bar{\Phi}_{\alpha }^{\,\,i}=-\delta _{\alpha }^{\,\,i\,\,}\varphi
^{\dagger }.$ Thus, 6 gauge parameters are used up in eliminating 6 degrees
of freedom from $\Phi .$ For a generic $\varphi ,$ the surviving symmetry is
a global diagonal sp$^{L+R}\left( 2,R\right) $ times a local noncommutative u%
$_{\star }^{L}\left( 1\right) \times u_{\star }^{R}\left( 1\right) .$ In
this gauge, the equations of motion become 
\begin{equation}
\mathcal{J}\star \varphi =-i\varphi \star V^{\prime }\left( -\varphi
^{\dagger }\star \varphi \right) ,\quad \mathcal{J}\star \mathcal{J}\mathbf{-%
}2i\mathcal{J}\mathbf{+}\left( \varphi \star \varphi ^{\dagger }\right) 1%
\mathbf{=}0.
\end{equation}
Rewriting the equations of motion in terms of components, we can separate
the triplet and singlet parts under the global sp$^{L+R}\left( 2,R\right) $ 
\begin{eqnarray}
J_{ij}\star \varphi &=&0,\quad J_{0}\star \varphi =\varphi \star V^{\prime
}\left( -\varphi ^{\dagger }\star \varphi \right) ,  \label{one} \\
\left( J_{0}+1\right) _{\star }^{2} &=&1+\varphi \star \varphi ^{\dagger }-%
\frac{1}{2}J_{ij}\star J^{ij}  \label{two} \\
\frac{1}{2}J_{(i}^{\,\,k}\star J_{j)k} &=&iJ_{ij}\star \left( J_{0}+1\right)
+i\left( J_{0}+1\right) \star J_{ij}  \label{three}
\end{eqnarray}
More explicitly, in terms of components, (using $J_{i}^{\,\,1}=J_{i2},$ and $%
J_{i}^{\,\,2}=-J_{i1}$) the left hand side of Eq.(\ref{three}) reduces to
commutators $\left[ J_{11},J_{12}\right] _{\star },$ $\left[ J_{11},J_{22}%
\right] _{\star }$ and $\left[ J_{11},J_{22}\right] _{\star }.$ In fact, if
in Eq.(\ref{three}) $J_{0}$ on the right hand side were absent, then the
commutation relations among the $J_{ij}$ would be precisely those of sp$%
\left( 2,R\right) $ as given in Eq.(\ref{sp2}). Then we should consider a
close relationship between $J_{ij}$ and $Q_{ij}.$ This is also supported by
the resemblance of Eq.(\ref{one}) to Eq.(\ref{matter}). Indeed, as we will
see below, the relation is nontrivial and interesting.

The second equation in (\ref{one}) can be written in the form $\left(
J_{0}+1\right) \star \varphi =\varphi \star \left( 1+V^{\prime }\left(
-\varphi ^{\dagger }\star \varphi \right) \right) .$ Applying $\left(
J_{0}+1\right) $ on both sides, using $\left( J_{0}+1\right) _{\star }^{2}$
given by Eq.(\ref{two}), and applying $J_{ij}\star \varphi =0$ as in Eq.(\ref
{one}), we obtain an equation purely for $\varphi $ 
\begin{equation}
\varphi \star \left( \left( 1+V^{\prime }\left( -\varphi ^{\dagger }\star
\varphi \right) \right) _{\star }^{2}-1-\varphi ^{\dagger }\star \varphi
\right) =0.  \label{scalar}
\end{equation}
It is straightforward to find all the solutions of this equation. Thus
consider any $\varphi =\lambda \varphi _{n}^{\,\,m},$ where $\lambda $ is a
complex constant, and $\varphi _{n}^{\,\,m}\left( X_{1},X_{2}\right) $ is of
the form of Eq.(\ref{wignerr}) which satisfies the triple relation of Eq.(%
\ref{tripple}) by construction. Then 
\begin{equation}
\varphi \star \varphi ^{\dagger }\star \varphi =\left| \lambda \right|
^{2}\varphi .
\end{equation}
Inserting such a $\varphi $ in the equation shows that $\lambda $ must be a
solution of the equation 
\begin{equation}
\left( 1+V^{\prime }\left( -\left| \lambda \right| ^{2}\right) \right)
^{2}-1-\left| \lambda \right| ^{2}=0.  \label{lambda}
\end{equation}
As an illustration, consider the example of the potential in Eq.(\ref{V}),
for which $V^{\prime }=a$ is a constant. For this case we find $\lambda =\pm
\left( a^{2}+2a\right) ^{1/2}.$ It is evident that, up to u$_{\star
}^{R}\left( 1\right) $ gauge transformations, the solutions of Eqs.(\ref{one}%
,\ref{scalar}) are all physical states of the form (\ref{wignerr})
multiplied by any $\left| \lambda \right| $ that solves the equation (the
phase of $\lambda $ can be absorbed away with a u$_{\star }^{R}\left(
1\right) $ transformation).

Next we solve $J_{0}$ formally from Eq.(\ref{two}), $J_{0}=-1+\left(
1+\varphi \star \varphi ^{\dagger }-\frac{1}{2}J_{ij}\star J^{ij}\right)
^{1/2},$ where the square root is understood as a power series with all
products replaced by star products. Using $J_{ij}\star \varphi =0=$ $\varphi
^{\dagger }\star J_{ij}$ and $\varphi \star \varphi ^{\dagger }\star \varphi
=\left| \lambda \right| ^{2}\varphi ,$ we can simplify each term in the
series expansion and obtain the simplified expression 
\begin{equation}
J_{0}=-1+\frac{V^{\prime }\left( -\left| \lambda \right| ^{2}\right) }{%
\left| \lambda \right| ^{2}}\varphi \star \varphi ^{\dagger }+\left( 1-\frac{%
1}{2}J_{ij}\star J^{ij}\right) ^{1/2}.  \label{Jo}
\end{equation}
where Eq.(\ref{lambda}) has been used. With this form of $J_{0},$ all
equations involving it, including the last one in Eq.(\ref{one}), are
satisfied. Finally, replacing these results into Eq.(\ref{three}) and using
again $J_{ij}\star \varphi =0=\varphi ^{\dagger }\star J_{ij},$ we find an
equation involving only the gauge fields $J_{ij},$ which we write in
components explicitly 
\begin{eqnarray}
\left[ J_{11},J_{12}\right] _{\star } &=&i\left\{ J_{11},\left(
1-C_{2}\left( J\right) \right) ^{1/2}\right\} _{\star }  \label{s1} \\
\left[ J_{11},J_{22}\right] _{\star } &=&2i\left\{ J_{12},\left(
1-C_{2}\left( J\right) \right) ^{1/2}\right\} _{\star }  \label{s2} \\
\left[ J_{12},J_{22}\right] _{\star } &=&i\left\{ J_{22},\left(
1-C_{2}\left( J\right) \right) ^{1/2}\right\} _{\star }  \label{s3}
\end{eqnarray}
The right hand side is a star anti-commutator involving the expression 
\begin{equation}
C_{2}\left( J\right) =\frac{1}{2}J_{kl}\star J^{kl}=\frac{1}{2}J_{11}\star
J_{22}+\frac{1}{2}J_{22}\star J_{11}-J_{12}\star J_{12}
\end{equation}
which looks like a Casimir operator. However, since Eqs.(\ref{s1}-\ref{s3})
are not the sp$\left( 2,R\right) $ Lie algebra one cannot hastily claim that 
$C_{2}\left( J\right) $ is a Casimir operator. Indeed, if one attempts to
derive the commutation relations between $C_{2}\left( J\right) $ and $J_{ij}$
by repeated use of Eqs.(\ref{s1}-\ref{s3}), one finds that $\left[
J_{ij},C_{2}\left( J\right) \right] _{\star }$ becomes equal to $\left[
-J_{ij},\left( \left( 1-C_{2}\left( J\right) \right) ^{1/2}\right) _{\star
}^{2}\right] $, and thus obtains an identity. Therefore, these equations do
not require that $C_{2}\left( J\right) $ and $J_{ij}$ commute. If they
commute, one could renormalize $J_{ij}$ by an appropriate factor to reduce
these equations to sp$\left( 2,R\right) $ commutation relations with the
normalization of generators as given by Eq.(\ref{sp2}). This is quite
interesting, as we will see below. Thus, generally Eqs.(\ref{s1}-\ref{s3})
are not the sp$\left( 2,R\right) $ commutation rules.

Furthermore, if one computes the Jacobi identities by repeated use of Eqs.(%
\ref{s1}-\ref{s3}), one finds 
\begin{equation}
\left[ J_{11},\left[ J_{12},J_{22}\right] _{\star }\right] _{\star }+cyclic=%
\frac{1}{2}\left[ J_{ij},\left[ J^{ij},\left( 1-C_{2}\left( J\right) \right)
^{1/2}\right] _{\star }\right] _{\star }.  \label{assoc}
\end{equation}
Under the assumption that the star product is associative, the Jacobi
identity is satisfied\footnote{%
It is also interesting to keep in mind the possibility of anomalies, leading
to non-associativity (e.g. magnetic fields \cite{zumino}). If we consider a
nonvanishing Jacobian, the mathematical structure of Eqs.(\ref{s1}-\ref{s3})
would be a Malchev algebra rather than a Lie algebra.}, and the left side of
Eq.(\ref{assoc}) vanishes. Therefore associativity of the star product
requires the right hand side to vanish, but generally this is a weaker
condition than the vanishing of $\left[ J_{ij},C_{2}\left( J\right) \right]
_{\star }.$

To solve the nonlinear gauge field equations (\ref{s1}-\ref{s3}) we will
setup a perturbative expansion around a background solution 
\begin{equation}
J_{ij}=J_{ij}^{\left( 0\right) }+gJ_{ij}^{\left( 1\right)
}+g^{2}J_{ij}^{\left( 2\right) }+\cdots
\end{equation}
such that $J_{ij}^{\left( 0\right) }$ is an exact solution and then analyze
the full equation perturbatively in powers of $g.$ For the exact background
solution we assume that $\frac{1}{2}J_{kl}^{\left( 0\right) }\star J^{(0)kl}$
commutes with $J_{ij}^{\left( 0\right) },$ therefore the background solution
satisfies a Lie algebra. Then we can write the exact background solution to
Eqs.(\ref{s1}-\ref{s3}) in the form 
\begin{equation}
J_{ij}^{\left( 0\right) }=Q_{ij}\star \frac{1}{\sqrt{1+\frac{1}{2}%
Q_{kl}\star Q^{kl}}}  \label{JQ}
\end{equation}
where $Q_{ij}$ satisfies the sp$\left( 2,R\right) $ algebra of Eq.(\ref{sp2}%
) 
\begin{equation}
\left[ Q_{11},Q_{12}\right] _{\star }=2iQ_{11},\quad \left[ Q_{11},Q_{22}%
\right] _{\star }=4iQ_{12},\quad \left[ Q_{12},Q_{22}\right] _{\star
}=2iQ_{22},  \label{Qalg}
\end{equation}
and $\frac{1}{2}Q_{kl}\star Q^{kl}$ is a Casimir operator that commutes with
all $Q_{ij}$ that satisfies the sp$\left( 2,R\right) $ algebra. The square
root is understood as a power series involving the star products and can be
multiplied on either side of $Q_{ij}$ since it commutes with the Casimir
operator. For such a background, the matter field equations (\ref{one})
reduce to 
\begin{equation}
Q_{ij}\star \varphi =0.  \label{Qphi}
\end{equation}
This is the matter field equation (\ref{matter}) given by the first
quantized theory. Its solution was discussed following Eq.(\ref{matter}).

Summarizing, we have shown that our action $S_{J,\Phi }$ has yielded
precisely what we had hoped for. The linearized equations of motion (0$^{th}$
power in $g$) in Eqs.(\ref{Qalg},\ref{Qphi}) are exactly those required by
the first quantization of the worldline theory as given by Eqs.(\ref{sp2},%
\ref{matter}). There remains to understand the propagation and self
interactions of the fluctuations of the gauge fields $gJ_{ij}^{\left(
1\right) }+g^{2}J_{ij}^{\left( 2\right) }+\cdots $, which are not included
in Eqs.(\ref{Qalg},\ref{Qphi}). However, the full field theory, without
making the assumption that $C_{2}\left( J\right) $ and $J_{ij}$ commute,
includes all the information. In particular the expansion of Eqs.(\ref{s1}-%
\ref{s3}) around the background solution $J_{ij}^{\left( 0\right) }$ of Eq.(%
\ref{JQ}) should determine uniquely both the propagation and the
interactions of the fluctuations involving photons, gravitons, and high spin
fields.

\subsection{Explicit background solution}

We record the exact solution to the background gauge field and matter field
equations (\ref{Qalg},\ref{Qphi}), which were obtained in several stages in 
\cite{field2T}\cite{highspin}\cite{NCSp}. The solution is given by fixing a
gauge with respect to the u$_{\star }^{L}\left( 1\right) .$ First, we choose
the gauge $Q_{11}=X_{1}\cdot X_{1}$. There is remaining u$_{\star
}^{L}\left( 1\right) $ symmetry that satisfies $\left[ X_{1}^{2},\omega _{0}%
\right] _{\star }=0.$ Using the conditions imposed on $Q_{12}$ by the sp$%
\left( 2,R\right) $ conditions, one finds that the remaining symmetry is
sufficient to fix a gauge for $Q_{12}=X_{1}\cdot X_{2}.$ Thus, up to u$%
_{\star }^{L}\left( 1\right) $ gauge transformations $\omega _{0}\left(
X_{1},X_{2}\right) $, one can simplify $Q_{11},Q_{12}$ and take the most
general $Q_{22}$ as follows 
\begin{equation}
Q_{11}=X_{1}^{M}X_{1}^{N}\eta _{MN},\quad Q_{12}=X_{1}^{M}X_{2M},\quad
Q_{22}=G\left( X_{1},X_{2}\right)  \label{bkg1}
\end{equation}
where $\eta ^{MN}$ is the flat metric in $d+2$ dimensions, and the general
function $G\left( X_{1},X_{2}\right) $ is assumed to have a power expansion
in $X_{2}$ in some domain 
\begin{eqnarray}
G\left( X_{1},X_{2}\right) &=&G_{0}\left( X_{1}\right) +G_{2}^{MN}\left(
X_{1}\right) \left( X_{2}+A\left( X_{1}\right) \right) _{M}\left(
X_{2}+A\left( X_{1}\right) \right) _{N}  \nonumber \\
&&+\sum_{s=3}^{\infty }G_{s}^{M_{1}\cdots M_{s}}\left( X_{1}\right)
\,\,\left( X_{2}+A\left( X_{1}\right) \right) _{M_{1}}\cdots \left(
X_{2}+A\left( X_{1}\right) \right) _{M_{s}}.  \label{bkg3}
\end{eqnarray}
The configuration space fields have the following interpretation: $%
A_{M}\left( X_{1}\right) $ is the Maxwell gauge potential, $G_{0}\left(
X_{1}\right) $ is a scalar, $G_{2}^{MN}\left( X_{1}\right) =\eta
^{MN}+h_{2}^{MN}\left( X_{1}\right) $ is the gravitational metric, and the
symmetric tensors $\left( G_{s}\left( X_{1}\right) \right) ^{M_{1}\cdots
M_{s}}$ for $s\geq 3$ are high spin gauge fields\footnote{%
There is no $G_{1}^{M}\left( X_{1}\right) $ as the coefficient of the first
power of $X_{2}+A,$ because $A_{M}\left( X_{1}\right) $ is equivalent to
that degree of freedom, as can be seen by re-expanding $Q_{22}$ in powers of 
$X_{2}$ instead of $X_{2}+A.$ Note also in $Q_{12}$ we really have $X_{2}+A,$
but $A$ has dropped because we chose to work in the gauge $X_{1}\cdot A=0.$}%
. The sp$\left( 2,R\right) $ closure condition in Eq.(\ref{Qalg}) requires
these background fields to be orthogonal to $X_{1}^{M}$ and to be
homogeneous of degree $(s-2)$ 
\begin{eqnarray}
X_{1}\cdot \partial A_{M} &=&-A_{M}\qquad X_{1}\cdot \partial G_{s}=\left(
s-2\right) G_{s},\quad  \label{hologr} \\
X_{1}^{M}A_{M} &=&X_{1M_{1}}h_{2}^{M_{1}M_{2}}=X_{1M_{1}}G_{s}^{M_{1}\cdots
M_{s}}=0,  \label{hologr2}
\end{eqnarray}
The background fields $A,G_{0},G_{2},G_{s\geq 3}$ determine all other
background fields $\left( f_{ij}\left( X_{1}\right) \right) ^{M_{1}\cdots
M_{s}}$ up to u$_{\star }^{L}\left( 1\right) $ gauge transformations $\omega
_{0}\left( X_{1},X_{2}\right) .$ The full solution of the $d+2$ dimensional
equations (\ref{hologr}) is given in \cite{highspin} in terms of $d$
dimensional background fields for Maxwell$,$ dilaton, metric, and higher
spin fields$.$ Therefore Eqs.(\ref{Qalg}) holographically encapsulate all
possible \textit{off-shell} arbitrary $d$ dimensional background gauge
fields in a $d+2$ dimensional formalism. In the next section we will derive
the dynamical equations of motion for the small fluctuations around the
backgrounds.

The u$_{\star }^{L}\left( 1\right) $ symmetry of the type $\omega _{0}\left(
X_{1},X_{2}\right) $, when expanded in powers of $X_{2}+A\left( X_{1}\right)
,$ contains the configuration space gauge transformation parameters for all
of the gauge fields \cite{highspin}. 
\begin{eqnarray}
\omega _{0}\left( X_{1},X_{2}\right) &=&\varepsilon _{0}\left( X_{1}\right)
+\varepsilon _{1}^{M}\left( X_{1}\right) \left( X_{2}+A\left( X_{1}\right)
\right) _{M}  \label{remain1} \\
&&+\sum_{s=2}^{\infty }\varepsilon _{s}^{M_{1}\cdots M_{s}}\left(
X_{1}\right) \ \left( X_{2}+A\left( X_{1}\right) \right) _{M_{1}}\cdots
\left( X_{2}+A\left( X_{1}\right) \right) _{M_{s}}  \nonumber
\end{eqnarray}
Since $Q_{11},Q_{12}$ have been gauge fixed, the remaining part of u$_{\star
}^{L}\left( 1\right) $ gauge symmetry should not change the form of $Q_{11}$%
, $Q_{12},$ so any surviving gauge parameters $\omega _{0}$ should satisfy $%
\left[ X_{1}^{2},\omega _{0}\right] _{\star }=\left[ \left( X_{1}\cdot
X_{2}\right) ,\omega _{0}\right] _{\star }=0$; this requires the gauge
parameters $\varepsilon _{s\geq 0}^{M_{1}\cdots M_{s}}\left( X_{1}\right) $
to be homogeneous and orthogonal to $X_{1}^{M}$ 
\begin{equation}
X_{1}\cdot \partial \varepsilon _{s}=s\,\varepsilon _{s},\qquad
X_{1M_{1}}\varepsilon _{s}^{M_{1}\cdots M_{s}}=0.  \label{remain2}
\end{equation}
The gauge transformation law of the gauge fields $\delta A,\delta
G_{0},\delta G_{2},\delta G_{s\geq 3}$ is given by $\delta Q_{22}=i\left[
Q_{22},\omega _{0}\right] _{\star }.$ From this it is easy to see that $%
\varepsilon _{0}\left( X_{1}\right) $ is the gauge parameter for the Maxwell
field, $\varepsilon _{1}^{M}\left( X_{1}\right) $ is the infinitesimal
general coordinate reparametrizations of all tensor fields, and $\varepsilon
_{s\geq 2}^{M_{1}\cdots M_{s}}\left( X_{1}\right) $ are the gauge parameters
for the high spin fields $G_{s+1}^{M_{1}\cdots M_{s+1}}.$ The details of the
gauge transformations are given in \cite{highspin}. This shows that the
familiar configuration space gauge principles, Maxwell, Einstein, and high
spin, are unified in our approach as being a small part of the u$_{\star
}\left( 1,1\right) $ gauge symmetry. We will use this remaining u$_{\star
}^{L}\left( 1\right) $ gauge symmetry in the analysis of the equations of
motion for the small fluctuations of the gauge fields.

\section{Fluctuations and dynamics of gauge fields}

We are interested in analyzing the perturbative expansion of Eqs.(\ref{s1}-%
\ref{s3}) around any background solution. In particular, taking a hint from
the form of Eqs.(\ref{bkg1}-\ref{bkg3}), we will investigate fluctuations $%
h\left( X_{1},X_{2}\right) $ in the direction of $J_{22}$. More general
fluctuations could also be considered\footnote{%
The u$_{\star }^{L}\left( 1\right) $ symmetry was sufficient to gauge fix
not only $Q_{11}$ but also $Q_{12},$ because these had to obey the sp$\left(
2,R\right) $ algebra. However, the $J_{ij}$ obey a more general set of
equations and therefore it is not clear whether one could gauge away more
general fluctuations around the background. There is certainly the freedom
to take vanishing fluctuations in the direction of $J_{11},$ but it is not
clear whether fluctuations in the direction of $J_{12}$ can also be
eliminated by gauge choices. This remains to be investigated.}, but we will
limit the current discussion to fluctuations around the background fields we
have identified in the previous section. Those are fluctuations in the
direction of $J_{22}$. Thus, we consider replacing any solution of the
background fields $A,G_{0},G_{2},G_{s\geq 3}$ by adding the fluctuations $%
\left( A+g\delta ^{\left( 1\right) }A\right) $, $\left( G_{0}+g\delta
^{\left( 1\right) }G_{0}\right) $, $\left( G_{2}+g\delta ^{\left( 1\right)
}G_{2}\right) $, $\,\left( G_{s\geq 3}+g\delta ^{\left( 1\right) }G_{s\geq
3}\right) $ and then expanding to first order in $g$. We already know from
the form of Eqs.(\ref{bkg1}-\ref{bkg3}), and the gauge transformations
discussed above, that these fluctuations are directly related to gauge
fields for Maxwell, Einstein and high spin gauge symmetries. We wish to
analyze the perturbative expansion of Eqs.(\ref{s1}-\ref{s3}) in order to
determine the equations of motion for the fluctuations.

The $J_{ij}$ including the fluctuations takes the form 
\begin{eqnarray}
J_{11} &=&\frac{1}{\sqrt{1+C_{2}\left( Q\right) }}\star X_{1}^{2},\quad
J_{12}=\frac{1}{\sqrt{1+C_{2}\left( Q\right) }}\star \left( X_{1}\cdot
X_{2}\right)  \label{J11} \\
J_{22} &=&\frac{1}{\sqrt{1+C_{2}\left( Q\right) }}\star G+gh\left(
X_{1},X_{2}\right) +\cdots  \label{Jh}
\end{eqnarray}
where $h\left( X_{1},X_{2}\right) $ is a general function while $G\left(
X_{1},X_{2}\right) $ describes a specific set of background fields that
solve Eqs.(\ref{hologr}-\ref{hologr2}). In particular, $G=X_{2}^{2}$
corresponds to the free flat background. $C_{2}\left( Q\right) $ is the
quadratic Casimir operator for the sp$\left( 2,R\right) $ algebra of Eq.(\ref
{Qalg}) satisfied by the background, and is given by 
\begin{eqnarray}
C_{2}\left( Q\right) &=&\frac{1}{2}Q_{kl}\star Q^{kl}=\frac{1}{2}\left(
Q_{11}\star Q_{22}+Q_{22}\star Q_{11}\right) -Q_{12}\star Q_{12} \\
&=&\frac{1}{2}\left\{ X_{1}^{2},G\right\} _{\star }-\left( X_{1}\cdot
X_{2}\right) \star \left( X_{1}\cdot X_{2}\right)  \label{c2Q} \\
&=&X_{1}^{2}G-\frac{1}{4}\left( \frac{\partial }{\partial X_{2}}\right)
^{2}G-\left( X_{1}\cdot X_{2}\right) ^{2}-\frac{1}{4}\left( d+2\right)
\end{eqnarray}
where, in the last line all star products have been evaluated. As long as
the background fields satisfy the equations (\ref{hologr}-\ref{hologr2})
this $C_{2}\left( Q\right) $ commutes with the sp$\left( 2,R\right) $
generators $Q_{ij}=\left( X_{1}^{2},\left( X_{1}\cdot X_{2}\right) ,G\left(
X_{1},X_{2}\right) \right) .$ The gauge field fluctuations $gh\left(
X_{1},X_{2}\right) $ up to first order in $g$ will be treated perturbatively
in solving the non-linear equations (\ref{s1}-\ref{s3}). Unlike the case of $%
J_{ij}^{\left( 0\right) },$ in this analysis we will not assume that $%
C_{2}\left( J\right) $ commutes with $J_{ij},$ and instead we will derive
the conditions that $h\left( X_{1},X_{2}\right) $ must satisfy to solve the
equations to first order in $g.$ We will develop the equations in the
general background $G\left( X_{1},X_{2}\right) $ up to a point and
eventually, for simplicity, specialize to the free background $G\left(
X_{1},X_{2}\right) \rightarrow X_{2}^{2}.$

First we compute $1-C_{2}\left( J\right) $ for the $J_{ij}$ in Eqs.(\ref{J11}%
,\ref{Jh}) to first order in $g$ 
\begin{eqnarray}
1-C_{2}\left( J\right) &=&1-C_{2}\left( J^{\left( 0\right) }\right) -\frac{g%
}{2}\left( J_{11}^{0}\star h+h\star J_{11}^{0}\right) \\
&=&\frac{1}{1+C_{2}\left( Q\right) }-\frac{g}{2\sqrt{1+C_{2}\left( Q\right) }%
}\star X_{1}^{2}\star h-h\star X_{1}^{2}\star \frac{g}{2\sqrt{1+C_{2}\left(
Q\right) }}.
\end{eqnarray}
Next we compute the square root up to first order in $g.$ Because of the
orders of factors, this is a complicated expression. In order to get a quick
glimpse of the content of the equations, for simplicity, we will proceed
under the assumption 
\begin{equation}
\left[ C_{2}\left( Q\right) ,h\right] _{\star }=0.
\end{equation}
Then we get the simple expression 
\begin{equation}
\sqrt{1-C_{2}\left( J\right) }=\frac{1}{\sqrt{1+C_{2}\left( Q\right) }}-%
\frac{g}{4}\left\{ X_{1}^{2},h\right\} _{\star }+\cdots .
\end{equation}
We now insert $J_{ij}$ in Eqs.(\ref{s1}-\ref{s3}) and expand both sides up
to the first power in $g$. The zeroth order terms cancel thanks to the
properties of $J_{ij}^{0}$, and the first power gives the following
equations that must be obeyed by $h\left( X_{1},X_{2}\right) $ 
\begin{eqnarray}
-\frac{i}{4}\left\{ X_{1}^{2},\left\{ X_{1}^{2},h\right\} _{\star }\right\}
_{\star } &=&0  \label{h1} \\
-\frac{i}{2}\left\{ \left( X_{1}\cdot X_{2}\right) ,\left\{
X_{1}^{2},h\right\} _{\star }\right\} _{\star } &=&\left[ X_{1}^{2},h\right]
_{\star }  \label{h2} \\
-\frac{i}{4}\left\{ G,\left\{ X_{1}^{2},h\right\} _{\star }\right\} _{\star
} &=&\left[ \left( X_{1}\cdot X_{2}\right) ,h\right] _{\star }-2ih
\label{h3}
\end{eqnarray}
We can show that the equation $\left[ C_{2}\left( Q\right) ,h\right] _{\star
}=0,$ that was assumed in arriving at these expressions, has no additional
information beyond these equations. That is, if we use $C_{2}\left( Q\right) 
$ as given in Eq.(\ref{c2Q}) and evaluate the commutator by repeatedly using
Eqs.(\ref{h1}-\ref{h3}), we find that $\left[ C_{2}\left( Q\right) ,h\right]
_{\star }=0$ is identically satisfied. Hence, this is not an additional
equation to be taken into account.

To get a quick grasp of the nature of these equations, we will first make a
few quick observations by assuming that the right hand side is zero, and in
the next paragraph we will analyze them by lifting this assumption. Thus,
after inserting the information $\left[ X_{1}^{2},h\right] _{\star }=0,$ and 
$\left[ \left( X_{1}\cdot X_{2}\right) ,h\right] _{\star }=2ih,$ we can
derive from the left hand side that $X_{1}^{2}\star H=0=H\star X_{1}^{2},$
and $\left( X_{1}\cdot X_{2}\right) \star H=0=H\star \left( X_{1}\cdot
X_{2}\right) ,$ and $\left\{ G,H\right\} _{\star }=0,$ where we have defined 
$H=\frac{1}{2}\left\{ X_{1}^{2},h\right\} _{\star }=X_{1}^{2}h-\partial
^{2}h/\left( \partial X_{2}\right) ^{2}$ after evaluating the star products.
The equations satisfied by $H$ are similar to Eqs.(\ref{Qalg},\ref{Qphi})
satisfied by a scalar field in a general background, in particular if we
consider $H$ similar to $\varphi \star \varphi ^{\dagger }$. As we have
already learned, the solution is of the form $H\sim \sum |\psi ><\psi
^{\prime }|,$ where the $\left\{ |\psi >\right\} $ satisfy the Klein-Gordon
equation in configuration space.

These remarks provide a quick indication that the first order equations (\ref
{h1}-\ref{h3}) for $h,$ or equivalently for $H,$ represent Klein-Gordon type
equations for the fluctuations of the gauge fields. In particular, it is
worth emphasizing that we have seen a first indication that the original
action includes a kinetic term for the fluctuations, although the formalism
does not make this immediately apparent. This point will become clearer in
the component form discussed in the next paragraph.

Next, we proceed to investigate Eqs. (\ref{h1}-\ref{h3}) in component
formalism without the assumptions of the previous two paragraphs. However,
to make further progress we will specialize to the free background $%
G=X_{2}^{2}$ and take the following expansion in powers of $X_{2},$ thus
defining the spin components of the fluctuating gauge fields in
configuration space 
\begin{equation}
h=h_{0}\left( X_{1}\right) +h_{1}^{M}\left( X_{1}\right) X_{2M}+\left(
X_{1}\right) \,X_{2M}X_{2N}+\sum_{s=3}^{\infty }h_{s}^{M_{1}\cdots
M_{s}}\left( X_{1}\right) \left( X_{2M_{1}}\cdots X_{2M_{s}}\right) .
\label{h}
\end{equation}
Up to factors of $\left( 1+C_{2}\left( Q\right) \right) ^{1/2}$ we have
redefined $h_{s}^{M_{1}\cdots M_{s}}$ as the fluctuations for the gauge
fields\footnote{%
Recall that the power expansion of $G\left( X_{1},X_{2}\right) $ did not
have $G_{1}^{M}$ associated with the first power of $X_{2},$ since the
Maxwell field $A_{M}$ was introduced as an independent field instead of $%
G_{1}^{M}$. The fluctuations of the Maxwell field appear gauge covariantly
everywhere in the form $X_{2}+g\delta ^{\left( 1\right) }A.$ The various
powers of this expression need to be expanded in powers of $g$ to first
order. However, there is already one power of $g$ in front of $%
h_{s}^{M_{1}\cdots M_{s}}$ since it is itself a fluctuation. Therefore, all $%
g\delta ^{\left( 1\right) }A$ drop out to first order in $g$. However, $%
g\delta ^{\left( 1\right) }A$ also appears in covariantizing the zeroth
order quadratic term $Q_{22}\rightarrow \left( X_{2}+g\delta ^{\left(
1\right) }A\right) ^{2}.$ The expansion of this term gives rise to $%
h_{1}^{M}\sim \delta ^{\left( 1\right) }A$ up to factors. Similarly, up to
overall factors, $h_{s}^{M_{1}\cdots M_{s}}$ is proportional to the
fluctuation $\delta ^{\left( 1\right) }G_{s}^{M_{1}\cdots M_{s}}$.}. In
particular, up to some factors, $h_{1}^{M}\left( X_{1}\right) $ is the
fluctuation of the Maxwell field, and $h_{2}^{MN}$ is the fluctuation of the
gravitational metric. The equations of motion can now be written for the
components by evaluating the star products in Eqs.(\ref{h1}-\ref{h3}). This
is done in the Appendix. Thus, $\left\{ X_{1}^{2},\left\{
X_{1}^{2},h\right\} _{\star }\right\} _{\star }$ =0\ \ gives in component
form 
\begin{equation}
\left( X_{1}^{2}\right) ^{2}h_{s}^{M_{1}\cdots M_{s}}-\frac{\left(
s+2\right) \left( s+1\right) }{2}X_{1}^{2}\eta _{MN}\,h_{s+2}^{MNM_{1}\cdots
M_{s}}+\frac{\left( s+4\right) \left( s+3\right) }{4}\eta _{KL}\,\eta
_{MN}\,h_{s+4}^{KLMNM_{1}\cdots M_{s}}=0  \label{fluc1}
\end{equation}
Similarly $\left[ X_{1}^{2},h\right] _{\star }=-\frac{i}{2}\left\{ \left(
X_{1}\cdot X_{2}\right) ,\left\{ X_{1}^{2},h\right\} _{\star }\right\}
_{\star }$\ gives 
\begin{eqnarray}
2\left( s+1\right) X_{1N}h_{s+1}^{NM_{1}\cdots M_{s}} &=&-\frac{2}{s}%
X_{1}^{(M_{1}}\left( X_{1}^{2}h_{s-1}^{M_{2}\cdots M_{s})}-\frac{s\left(
s+1\right) }{4}\eta _{MN}h_{s+1}^{MNM_{2}\cdots M_{s})}\right)  \label{fluc2}
\\
&&-\left( s+1\right) \partial _{N}\left( X_{1}^{2}h_{s+1}^{NM_{1}\cdots
M_{s}}-\frac{\left( s+3\right) \left( s+2\right) }{4}\eta
_{KL}h_{s+3}^{KLNM_{1}\cdots M_{s}}\right)  \nonumber
\end{eqnarray}
and $\left[ \left( X_{1}\cdot X_{2}\right) ,h\right] _{\star }-2ih=-\frac{i}{%
4}\left\{ X_{2}^{2},\left\{ X_{1}^{2},h\right\} _{\star }\right\} _{\star }$%
\ gives 
\begin{eqnarray}
\left( s-2-X_{1}\cdot \partial _{1}\right) h_{s}^{M_{1}\cdots M_{s}} &=&%
\frac{-2}{s\left( s-1\right) }\eta ^{(M_{1}M_{2}}\left(
X_{1}^{2}h_{s-2}^{M_{3}\cdots M_{s})}-\frac{s\left( s-1\right) }{4}\eta
_{MN}h_{s}^{MNM_{3}\cdots M_{s})}\right)  \nonumber \\
&&+\frac{1}{4}\partial _{1}^{2}\left( X_{1}^{2}h_{s}^{M_{1}\cdots M_{s}}-%
\frac{\left( s+2\right) \left( s+1\right) }{4}\eta
_{MN}h_{s+2}^{MNM_{1}\cdots M_{s}}\right)  \label{fluc3}
\end{eqnarray}
These equations are purely in configuration space $X_{1}^{M}.$ The first two
equations may be interpreted as subsidiary conditions, while the last one is
a second order Klein-Gordon type equation. By construction, they are gauge
invariant under the remaining gauge transformations $\omega _{0}\left(
X_{1},X_{2}\right) .$ Since we have $\left[ C\left( Q\right) ,h\right] =0,$
the remaining gauge symmetry also obeys 
\begin{equation}
\left[ C\left( Q\right) ,\omega _{0}\right] =0
\end{equation}
in addition to Eq.(\ref{remain2}), hence they are a subset of the gauge
transformations discussed in \cite{highspin} These gauge transformations do
not change the form of $J_{11},J_{12},$ while they are applied to the total $%
J_{22}$ as $\delta J_{22}=i\left[ J_{22},\omega _{0}\right] _{\star }$ from
which the transformation properties for the components $h_{s}$ are obtained.

Note that the double trace of $h_{s\geq 4}$ is restricted by Eq.(\ref{fluc1}%
), an important fact for high spin gauge theories \cite{fronsdal}. In this
connection, we may ask if the double trace would vanish when the $d+2$
dimensional system is holographically viewed in $d$ dimensions. As part of
the reduction from $d+2$ dimensions to $d$ dimensions we need to impose the
vanishing of $X_{1}^{2}$. Although $X_{1}^{2}h_{s}^{M_{1}\cdots M_{s}}$ does
not vanish, it appears that $\left( X_{1}^{2}\right) ^{2}h_{s}^{M_{1}\cdots
M_{s}}$ and $\eta _{MN}X_{1}^{2}h_{s}^{MNM_{1}\cdots M_{s}}$ may
consistently be taken to vanish. Then at the end of the holographic
reduction the double trace does indeed vanish in $d$ dimensions.

The main point established in this section is that the full non-linear
equations contain information on the propagation of the gauge fields. For
simplicity, this was done under the assumption $\left[ C_{2}\left( Q\right)
,h\right] _{\star }=0.$ It is desirable to analyze the full form of the
perturbative expansion without relying on this assumption. Also, there still
remains the completion of this exercise to extract the full form of the
kinetic terms and interactions after the reduction to a holographic picture
in $d$ dimensions. At that point it will be interesting to compare our
equations for the high spin gauge fields to those discussed in other
formalisms. In previous investigations equations of motion have been
constructed in $d=3,4$ dimensions, including up to cubic interactions that
satisfy a truncated (or approximate) form of a high spin gauge symmetry \cite
{vasil}.  But the general interaction is not known, and furthermore the
construction of an off-shell action has eluded all attempts. By contrast,
our approach begins with a complete and unique action (modulo the cubic
condition). It is already clear that our theory supplies both the
propagation and all interactions of the gauge fields. It would be very
interesting to investigate the relation between our approach and that of 
\cite{vasil}. Related aspects of high spin gauge fields are still under
study in our theory, and we hope to report on this topic in a future
publication.

\section{Matrix point of view}

In some sense, our current noncommutative field theory is an infinite
dimensional matrix theory, and it can be viewed as the large $N$ limit of a
finite 2$N\times 2N$ matrix theory.

The fields $J_{ij}\left( X_{1},X_{2}\right) $ and $J_{0}\left(
X_{1},X_{2}\right) $ are constructed from noncommutative $d+2$ dimensional
phase space $\left( X_{1}^{M},X_{2M}\right) .$ Using the Weyl
correspondence, it is possible to replace $\left( X_{1}^{M},X_{2M}\right) $
by quantum operators acting in a Hilbert space, or equivalently by infinite
dimensional matrices. In this sense, our theory is already a ``matrix
theory'' for infinite dimensional matrices.

One can introduce a cutoff in the theory by replacing the matrices by finite
matrices. The basic Heisenberg commutation rules between $\left(
X_{1}^{M},X_{2M}\right) $ cannot be obeyed by finite matrices, but by taking
special combinations of the basic operators $\left( X_{1}^{M},X_{2M}\right) $
one can confine oneself to quantities $J_{ij}$ constructed from them, such
that $J_{ij}$ are finite matrices. For example, this is the case on a
periodic torus where finite translations in phase space $u_{a}=\exp \left(
ia\cdot X_{1}\right) $ and $v_{b}=$exp$\left( ib\cdot X_{2}\right) $ \ are
indeed represented by finite matrices that obey the algebra $%
u_{a}v_{b}=v_{b}u_{a}\omega _{ab}$ when $\omega _{ab}=\exp \left( -ia\cdot
b\right) $ is a root of unity. Similar considerations apply to the fuzzy
sphere in phase space (with $\left( d,2\right) $ signature in our case).

Therefore, it is possible to take $J_{ij}$ and $J_{0}$ as functions of only $%
u_{a},v_{b}$ (for a collection of $a$'s and $b$'s), or similar structures,
and thus represent them as functions of finite matrices that are closely
connected to phase space $\left( X_{1}^{M},X_{2M}\right) $. We expect then
the non-commutative u$_{\star }\left( 1,1\right) $ to be approximated by the
non-compact group u$\left( N,N\right) $ such that the $2\times 2$
noncommutative $\mathcal{J}$ \ gets replaced by the 2$N\times 2N$ matrix
representation of u$\left( N,N\right) .$ The four $N\times N$ blocks are
then identified with the hermitian $J_{ij},J_{0}$ just as in Eq.(\ref{J}).
The form of the action formally remains the same as Eq.(\ref{SZ}), except
for replacing integration by a trace over matrices. Thus, the equations that 
$\mathcal{J}$ satisfies are also formally the same, except for replacing
star products with matrix products.

We now face again the matrix analog of Eqs.(\ref{s1}-\ref{s3}), instead of
star products. When $\frac{1}{2}J_{kl}J^{kl}$ commutes with $J_{ij}$ it is
possible to construct $Q_{ij},Q_{0}$ that satisfy the u$\left( N,N\right) $
algebra, as in Eq.(\ref{JQ}). However, the solution for $Q_{ij},Q_{0}$ must
now be given in terms of $u_{a},v_{b}.$ Indeed it is possible to construct
the u$\left( N,N\right) $ algebra in terms of powers of $u_{a},v_{b}$ or
similar structures, just like the examples that exist in the literature for $%
U\left( 2N\right) $ \cite{(1)}-\cite{hoppe}. This would provide the matrix
analog of the background solution in Eqs.(\ref{bkg1}-\ref{bkg3}).

Since there are many solutions of the type Eqs.(\ref{bkg1}-\ref{bkg3}) we
expect also many solutions for $Q_{ij},Q_{0}$ as functions of $u_{a},v_{b}$
or some similar structures. The more general solution of Eqs.(\ref{s1}-\ref
{s3}) for $J$'s that include propagation of the gauge fields, can then be
investigated using finite matrix methods.

\section{Outlook}

We have learned that we can consistently formulate a field theory of
2T-physics in $d+2$ dimensions based on a very basic gauge principle in
quantum phase space. We have tentatively shown that our equations, compactly
written in phase space in the form of Eq.(\ref{generaleq}), seem to yield a
new unified description of various gauge fields in configuration space,
including Maxwell, Einstein, and high spin gauge fields interacting with
matter and among themselves in $d$ dimensions.

The underlying gauge principle is the noncommutative u$_{\star }\left(
1,1\right) ,$ and the action that gives rise to the field equations in
noncommutative phase space has the rather simple form of Eq.(\ref{SZ}). As
argued following Eq.(\ref{V}), the form of the action is unique as long as
it is restricted to the maximum cubic power of $J.$ Then, all results are
grounded purely in the u$_{\star }\left( 1,1\right) $ gauge principle$.$
With the only assumption being the cubic restriction, the worldline approach
is explained by the field theory as \textit{exact background solutions}.
This essentially unique action could now be taken as a starting point for a
classical as well as quantum analysis of the \textit{interacting} 2T-physics
field theory. At this time it is not known what would be the consequences of
relaxing the maximum cubic power of $J$.

Although the analysis of the classical field equations of motion so far has
been rudimentary, it was sufficient for showing that the content of the
theory is sensible while being very rich and interesting from the point of
view of $d$ dimensions. As usual, the 1T-physics content of the theory can
be obtained as various holographic images that come from embedding $d$
dimensions in various ways in $d+2$ dimensions. One of the better understood
holographic images \cite{Dirac}\cite{field2T} is the field theory in $d$
dimensions in which the Klein-Gordon matter field interacts with various
gauge fields, including interactions with the Maxwell field, dilaton,
gravitational field, and high spin gauge fields.

The gauge fields propagate and have interactions among themselves. It
appears that our approach provides for the first time an action principle
that should contribute to the resolution of the long studied but unfinished
problem of high spin fields \cite{vasil}\cite{segal}\cite{highspin}\cite
{sezgin}. We have shown that there is a kinetic term for the gauge fields
although more study is needed to understand its contents better. The nature
and detail of the interactions among the gauge fields can in principle be
extracted from our $d+2$ dimensional theory, but this remains as an exercise
for the future\footnote{%
The nature of interactions for the high spin fields may depend on the
background chosen for $Q_{ij}$. For example, according to previous studies 
\cite{vasil} there are no interactions in flat backgrounds, but there are
interactions in AdS$_{d}$ backgrounds, in particular $d=3,4$. In 2T physics,
flat backgrounds or AdS$_{d}$ backgrounds both exist in the same $d+2$
dimensional theory as they emerge from different embeddings of $d$
dimensions in $d+2$ dimensions (see last reference in \cite{old2T} and \cite
{field2T}). So it should be interesting to study such issues in detail with
regard to high spin field interactions.}.

This work can be generalized in several directions. One of these directions
is supersymmetry, and one can consider both worldline and space-time
supersymmetries.

In the case of worldline supersymmetry, local sp$\left( 2,R\right) $ is
replaced by local osp$\left( n|2\right) $ where $n$ is the number of
supersymmetries. This describes 2T-physics for spinning particles \cite
{spin2t}. Local osp$\left( n|2\right) $ on the worldline can be maintained
in the presence of background fields, and this has been studied to some
extent \cite{field2T}, but more work along the present paper, to build a
noncommutative field theory, remains to be done. We may guess that the
appropriate gauge group for the supersymmetric noncommutative field theory
would be u$_{\star }\left( n|1,1\right) .$ Therefore it would be interesting
to take the same form of the action in Eq.(\ref{SZ}) and repeat the analysis
of the current paper for the noncommutative supergroup u$_{\star }\left(
n|1,1\right) .$ It is likely that the content of this theory is the spinning
generalization of what we have discussed in this paper.

In the case of space-time supersymmetry, the worldline action in the absence
of background fields has been constructed \cite{super2t}\cite{survey2T}. The
local symmetries are richer: in addition to local sp$\left( 2,R\right) $
they include a $d+2$ dimensional version of kappa supersymmetry and its
bosonic generalizations. For the special supersymmetries osp$\left(
N|4\right) ,$ su$\left( 2,2|N\right) ,$ F$\left( 4\right) $ and osp$\left(
6,2|N\right) $ one obtains a $d+2$ dimensional formulation of the
superconformal particle in $d=3,4,5,6$ dimensions respectively. For other
supergroups one obtains brane collective coordinates in interaction with
superparticle coordinates, giving unitary supersymmetric BPS states as the
quantum states of the theory. In particular, for osp$(1|64)$ one obtains a
toy M-model that embodies certain interesting features of M-theory \cite
{super2t}\cite{survey2T}.

The case of background fields in the presence of space-time supersymmetry in
the worldline theory remains to be constructed. We expect this to be a
rather interesting and rewarding exercise, because kappa supersymmetry is
bound to require the background fields to satisfy dynamical equations of
motion, as it does in 1T physics \cite{wittsusy}. The supersymmetric field
equations thus obtained in $d+2$ dimensions should be rather interesting as
they would include some long sought field theories in $d+2$ dimensions,
among them super Yang-Mills and supergravity theories. Perhaps one may also
attempt directly the space-time supersymmetrization of the present approach,
bypassing the background field formulation of the worldline theory.

The matrix approach described above should eventually be considered with
space-time supersymmetry. It is conceivable that these methods would lead to
a formulation of covariant M(atrix) theory. In this context we expect osp$%
\left( 1|64\right) $ to play a crucial role, as some of its attractive
features appear to be quite relevant to M-theory \cite{S-theory}\cite
{super2t}\cite{west}\cite{ferrara}. In future work we intend to pursue the
types of issues that are touched upon in this section.

\section*{Acknowledgments}

The last stages of this work were completed while I was a participant in the
M-theory program at the ITP at Santa Barbara. I gratefully acknowledge the
support of the ITP.

I would like to thank C. Deliduman, D. Minic, S.J. Rey, M. Vasiliev, M.M.
Sheikh-Jabbari and E. Witten for discussions and comments.

\section{Appendix}

Let $h\left( X_{1},X_{2}\right) $ be given by the expansion in Eq.(\ref{h}).
We compute explicitly 
\begin{eqnarray}
&=&X_{1}^{2}h-\frac{1}{4}\partial _{2}^{2}h=H \\
&=&\sum_{s=0}^{\infty }\left( X_{1}^{2}h_{s}^{M_{1}\cdots M_{s}}-\frac{%
\left( s+2\right) \left( s+1\right) }{4}\eta _{MN}h_{s+2}^{MNM_{1}\cdots
M_{s}}\right) \left( X_{2M_{1}}\cdots X_{2M_{s}}\right)
\end{eqnarray}
By applying the formula a second time we compute $\frac{1}{2}\left\{
X_{1}^{2},H\right\} _{\star }.$ Then the component form of Eq.(\ref{h1})
gives Eq.(\ref{fluc1}).

Next we compute the commutator 
\begin{eqnarray}
\left[ X_{1}^{2},h\right] _{\star } &=&2iX_{1}\cdot \partial _{2}h \\
&=&2i\sum_{s=0}^{\infty }\left( s+1\right) X_{1N}h_{s+1}^{NM_{1}\cdots
M_{s}}\,\left( X_{2M_{1}}\cdots X_{2M_{s}}\right)
\end{eqnarray}
and anticommutator $\ $%
\begin{eqnarray}
\frac{1}{2}\left\{ \left( X_{1}\cdot X_{2}\right) ,H\right\} _{\star }
&=&\left( X_{1}\cdot X_{2}\right) H+\frac{1}{4}\left( \partial _{1}\cdot
\partial _{2}\right) H \\
&=&\sum_{s=0}^{\infty }\left( \frac{1}{s}X_{1}^{(M_{1}}H_{s-1}^{M_{2}\cdots
M_{s})}+\frac{\left( s+1\right) }{4}\partial _{N}H_{s+1}^{NM_{1}\cdots
M_{s}}\right) \left( X_{2M_{1}}\cdots X_{2M_{s}}\right)
\end{eqnarray}
We use them in computing the component form of Eq.(\ref{h2}), which gives
Eq.(\ref{fluc2})

Finally we compute the commutator 
\begin{eqnarray}
\left[ \left( X_{1}\cdot X_{2}\right) ,h\right] _{\star } &=&i\left(
X_{2}\cdot \partial _{2}-X_{1}\cdot \partial _{1}\right) h \\
&=&i\sum_{s=0}^{\infty }\left( \left( s-X_{1}\cdot \partial _{1}\right)
h_{s}^{M_{1}\cdots M_{s}}\right) \left( X_{2M_{1}}\cdots X_{2M_{s}}\right)
\end{eqnarray}
and anticommutator 
\begin{eqnarray}
\frac{1}{2}\left\{ X_{2}^{2},H\right\} _{\star } &=&X_{2}^{2}H-\frac{1}{4}%
\partial _{1}^{2}H \\
&=&\sum_{s=0}^{\infty }\left( \frac{2}{s\left( s-1\right) }\eta
^{(M_{1}M_{2}}H_{s-2}^{M_{3}\cdots M_{s})}-\frac{1}{4}\partial
_{1}^{2}H_{s}^{M_{1}\cdots M_{s}}\right) \left( X_{2M_{1}}\cdots
X_{2M_{s}}\right)
\end{eqnarray}
By inserting them in Eq.(\ref{h3}) we obtain the component form given in Eq.(%
\ref{fluc3}).


\begin{thebibliography}{99}
\bibitem{survey2T}  I. Bars, \textit{``}Survey of Two-Time Physics\textit{''}%
, hep-th/0008164.

\bibitem{old2T}  I. Bars, C. Deliduman and O. Andreev, Phys. Rev. \textbf{D58%
} \ (1998) 066004, hep-th/9803188; I. Bars, Phys. Rev. \textbf{D58} (1998)
066006, hep-th/9804028; S. Vongehr, hep-th/9907077; I. Bars, hep-th/9809034;
I. Bars, Phys. Rev. \textbf{D59} (1999) 045019, hep-th/9810025;

\bibitem{spin2t}  I. Bars and C. Deliduman, Phys. Rev. \textbf{D58} (1998)
106004, hep-th/9806085;

\bibitem{super2t}  I. Bars, C. Deliduman and D. Minic, ``Supersymmetric
Two-Time Physics\textit{'', }Phys. Rev. \textbf{D59} (1999) 125004,
hep-th/9812161; ``Lifting M-theory to Two-Time Physics\textit{'', }Phys.
Lett. \textbf{B457} (1999) 275, hep-th/9904063. \newline
I. Bars, ``\textit{2}T formulation of superconformal dynamics relating to
twistors and supertwistors'', Phys. Lett. \textbf{B483} (2000) 248,
hep-th/0004090; ``\textit{AdS}$_{5}\times $\textit{\ S}$^{5}$\textit{\ }%
Supersymmetric Kaluza-Klein Towers as a 12-dimensional Theory in
2T-Physics'', in preparation (partial results in \cite{survey2T}); ``A Toy
M-model'', in preparation (partial results in hep-th/9904063 and \cite
{survey2T}).

\bibitem{string2t}  I. Bars, C. Deliduman and D. Minic, ``String, Branes and
Two-Time Physics\textit{'', }Phys. Lett. \textbf{B466} (1999) 135,
hep-th/9906223.

\bibitem{emgrav}  I. Bars, ``Two-Time Physics with Gravitational and Gauge
Field Backgrounds\textit{'', }Phys. Rev. \textbf{D62} (2000) 085015,
hep-th/0002140.

\bibitem{field2T}  I. Bars, ``Two-Time Physics in Field Theory\textit{'', }%
Phys. Rev. \textbf{D62} (2000) 046007, hep-th/0003100.

\bibitem{highspin}  I. Bars and C. Deliduman,\ ``High spin gauge fields and
Two-Time Physics'', hep-th/0103042.

\bibitem{NCSp}  I. Bars and S.J. Rey, ``Noncommutative Sp(2,R) Gauge
Theories - A Field Theory Approach to Two-Time Physics'', hep-th/0104135.

\bibitem{ncOn}  I. Bars, M.M. Sheikh-Jabbari, M.A. Vasiliev,
``Noncommutative o*(N) and usp*(2N) algebras and the corresponding gauge
field theories'', hep-th/0103209.

\bibitem{weyl}  H. Weyl, Zeit. f\"ur Phys. \textbf{46} (1927) 1.

\bibitem{wigner}  E. Wigner, Phys. Rev. \textbf{40} (1932) 749; in\textit{\
Perspectives in Quantum Theory}, Eds. W. Yourgrau and A. van de Merwe (MIT
Press, Cambridge, 1971).

\bibitem{moyal}  J. Moyal, Proc. Camb. Phil. Soc. \textbf{45 }(1949)\textbf{%
\ }99.

\bibitem{Dirac}  P.A.M Dirac, Ann. Math. 37 (1936) 429; H. A. Kastrup, Phys.
Rev. 150 (1966) 1183; G. Mack and A. Salam, Ann. Phys. 53 (1969) 174; C. R.
Preitschopf and M. A. Vasiliev, hep-th/9812113.

\bibitem{fronsdal}  C. Fronsdal, ``Massless Fields With Integer Spin'',
Phys. Rev. \textbf{D18} (1978) 3624.

\bibitem{vasil}  M. A. Vasiliev, Int. J. Mod. Phys. \textbf{D5 }(1996) 763;
``Higher spin gauge theories, star products and AdS space'', hep-th/9910096.

\bibitem{segal}  A. Yu. Segal, ``Point particle in general background fields
and generalized equivalence principle'', hep-th/0008105; ``A Generating
formulation for free higher spin massless fields'', hep-th/0103028.

\bibitem{sezgin}  E. Sezgin and P. Sundell, ``Doubletons and higher spin
gauge theories'', hep-th/0105001.

\bibitem{zumino}  M. G\"{u}naydin and B. Zumino, ``Magnetic Charge And
Non-associative Algebras'', LBL-19200, in Pisa 1984 Proceedings\textit{, New
and Old Problems in Fundamental Physics.}

\bibitem{(1)}  J. Goldstone, unpublished; J. Hoppe, Int. J. Mod. Phys., 
\textbf{A4} (1989) 5235; J. Hoppe, MIT Ph.D. Thesis, Elem. Part. Res. J.
(Kyoto) \textbf{80} (89/90) no.3.

\bibitem{(2)}  D. B. Fairlie, P. Fletcher and C.K. Zachos, Phys. Lett. 
\textbf{B218} (1989) 203. D.B. Fairlie and C.K. Zachos, Phys. Lett. \textbf{%
B224} (1990) 101; V. Arnold Ann. Inst. Fourier XVI, No.\textbf{1} (1966) 319.

\bibitem{(3)}  E. Floratos and J. Illiopoulos, Phys. Lett. \textbf{B201}
(1988) 237; E. Floratos, Phys. Lett. \textbf{B228} (1989) 335. E. Floratos
and S. Nicolis, J. Phys. A31 (1998) 3961.

\bibitem{ibmatrix}  I. Bars, Phys. Lett. \textbf{B245} (1990) 35; ``Strings
and matrix models on genus $g$ Riemann surfaces'', hep-th/9706177.

\bibitem{(4)}  C. N. Pope and L. Romans, Class. Quantum Grav. \textbf{7}
(1990) 97; D. B. Fairlie and C. K. Zachos, J. Math. Phys. \textbf{31} (1990)
1088.

\bibitem{(10)}  I. Bars, C. Pope and E. Sezgin, Phys. Lett. \textbf{B210}
(1988) 85.

\bibitem{mem}  B. de Wit, J. Hoppe and H. Nicolai, Nucl. Phys. \textbf{B305}
(1988) 545.

\bibitem{cor}  L. Cornalba and R. Schiappa, ``Matrix Theory Star Products
from the Born-Infeld Action", hep-th/9907211.

\bibitem{(9)}  G. 'tHooft, Comm. Math. Phys. \textbf{81} (1981) 267.

\bibitem{hoppe}  J. Hoppe, Phys. Lett. \textbf{B215} (1988) 706; A. O.
Morris, J. London Math. Soc. (2) \textbf{7} (1973) 235.

\bibitem{BFSS}  T. Banks, W. Fischler, S.H. Shenker and L. Susskind, Phys.
Rev. \textbf{D55} (1997) 5112, hep-th/9610043. \newline
N. Ishibashi, H. Kawai, Y. Kitazawa and A. Tsuchiya, Nucl. Phys. \textbf{B498%
} (1997) 467, hep-th/9612115.

\bibitem{wittsusy}  E. Witten, ``Twistor-like transform in ten-dimensions'',
Nucl.Phys.B266:245,1986.

\bibitem{S-theory}  I. Bars, ``S-theory'', Phys. Rev. \textbf{D55} (1997)
2373, [hep-th/9607112]; ``Algebraic structure of S theory'', hep-th/9608061.

\bibitem{west}  P. West, ``Hidden superconformal symmetry in M theory'', 
\textbf{JHEP} (2000) 0008:007, [hep-th/0005270].

\bibitem{ferrara}  R. D'Auria, S. Ferrara and M.A. Lledo, ``On the embedding
of space-time symmetries into simple superalgebras'', hep-th/0102060.
\end{thebibliography}
\end{document}